\documentclass{article}
\usepackage{amsmath, amssymb}
\usepackage[dvipdfmx]{graphicx}
\newtheorem{theo}{Theorem}[section]

\newtheorem{lemm}[theo]{Lemma}
\newtheorem{rem}[theo]{Remark}
\def\proof{\textbf{Proof}}

\begin{document}
\title{\textbf{Hypergeometric Solutions for the $q$-Painlev\'{e} Equation of type $E^{(1)}_6$ by Pad\'{e} Method}}
\author{Yusuke Ikawa\\
{\small Department of Mathematics, Faculty of Science, Kobe University, Hyogo 657-8501, Japan}\\
{\small Email: yikawa@math.kobe-u.ac.jp}\\
}
\date{}
\maketitle
\noindent Abstract: The $q$-Painlev\'{e} equation of  type $E^{(1)}_6$ is obtained by Pad\'{e} method. Special solutions in determinant formula to the $q$-Painlev\'{e} equation is presented. A relation between Pad\'{e} method and B\"{a}cklund transformation of type $E^{(1)}_6$ is given.
\section{Introduction}
$q$-Painlev\'{e} equations and their solutions have been studied from various viewpoints. Especially, Sakai gave a natural classification of discrete Painlev\'{e} equations by means of the geometry of rational surfaces \cite{sakai}. 
In this classification, discrete Painlev\'{e} equations are classified by means of affine Weyl groups.

For the solutions of $q$-Painlev\'{e} equations, seed solutions of hypergeometric type were obtained (see for instance \cite{kmnoy}).
More general solutions obtained by applying the B\"{a}cklund transformations have been given in \cite{hk}\cite{hk2}\cite{kajiwara31}\cite{kajiwara32}\cite{kajiwara4}\cite{masuda}\cite{masuda2}\cite{sakai6}.
These results are based on the bilinear relations for the $\tau$-functions.

$q$-Painlev\'{e} equation of type $E^{(1)}_6$, a main subject in this paper is first proposed in \cite{RGH}.
We study the equation in the form appeared in \cite{RGO} i.e. the equations (\ref{painlevebasis}) and (\ref{painlevebasis2}) where $f_1$ and $g_1$ are variables and $b_1, \cdots, b_8$ are parameters.
And in \cite{kmnoy}, a hypergeometric solution to the $q$-Painlev\'{e} equations is expressed in terms of the $q$-hypergeometric series (\ref{qhgs}). 
Though the seed solution (\ref{qhgs}) for the $q$-Painlev\'{e} equation is already known, any results have not been obtained for its B\"{a}cklund transformed solutions so far.

In this paper, we will construct the hypergeometric solutions to the $q$-Painlev\'{e} equation of type $E^{(1)}_6$ in a determinant formula 
by using Pad\'{e} method \cite{yamada}\cite{yoshioka}, 
which is one method to derive Lax equations to the Painlev\'{e} equation. 
Applying this method to $q$-Painlev\'{e} equations were not accomplished for $E^{(1)}_6$ case before.

This paper is organized as follows. In section 2, we will get $q$-Painlev\'{e} VI equation\cite{jimbosakai} by Pad\'{e} method (Theorem\ref{ccqp6}) as warming up example. 
Using this method, we will obtain $q$-Painlev\'{e} equation of type $E^{(1)}_6$ (Theorem\ref{cce6})  and its explicit solutions in determinant formula (Theorem \ref{theoremfandg}) in section 3. 
In appendix \ref{qrt}, we will discuss the relation between QRT system\cite{QRT} and $q$-Painlev\'{e} equations. 
Finally in appendix \ref{backlund}, we will study a relation between the Pad\'{e} method and the B\"{a}cklund transformation.

\section{The case of $D^{(1)}_5$ ($q$-Painlev\'{e}VI)}\label{secqp6}

In this section, we will derive the Lax pair for $D^{(1)}_5$ $q$-Painlev\'{e} equation (Theorem \ref{subseccde}) and the $q$-Painlev\'{e} VI equation (Theorem \ref{subseccc}) by using the Pad\'{e} method.

\subsection{Construction of the difference equation}\label{subseccde}

Let $a_1, a_2, a_3, a_4 \in \mathbb{C}\backslash \{0\}$ be complex parameters. Define a function $Y(x)$ as
\begin{equation}
Y(x)=\frac{(a_1x,a_2x)_{\infty}}{(a_3x,a_4x)_{\infty}}
\end{equation}
where the symbol $(\quad)_i$ is defined as
\begin{equation}
(z_1, z_2, \cdots, z_i)_j = \prod_{k=0}^{j}(1-q^kz_1)(1-q^kz_2)\cdots(1-q^kz_i)\nonumber.
\end{equation}

Let $m, n \in \mathbb{Z}_{\geq 0}$ and put $N=m+n$. We consider  the approximation  of $Y(x)$ by rational function

\begin{equation}\label{padekinji}
Y(x)\equiv \frac{P_m(x)}{Q_n(x)}\quad(\mathrm{mod} \; x^{N+1})
\end{equation}
where $P_m(x)$ and $Q_n(x)$ are polynomials of order $m$ and $n$ respectively and  the constant term of $Q_n(x)$ is $1$.

We will construct two difference equations whose solutions are $P_m(x)$ and $Y(x)Q_n(x)$.

We define a transformation $T$ as
\begin{equation}
T: a_2\mapsto q a_2,\quad a_4\mapsto qa_4\nonumber
\end{equation}
and we denote as $T(y)=\overline{y}$ and $T^{-1}(y)=\underline{y}$. Here and in the following, we will write only non-trivial actions. Using this transformation $T$, we will construct following difference equations
\begin{equation}\label{l2}
L_2:
\begin{array}{|ccc|}
y(x) & y(qx) & \overline{y}(x) \\
P_m(x) & P_m(qx) & \overline{P_m}(x) \\
Y(x)Q_n(x) & Y(qx)Q_n(qx) & \overline{Y}(x)\overline{Q_n}(x)
\end{array}
=0 ,
\end{equation}
\begin{equation}\label{l3}
L_3: 
\begin{array}{|ccc|}
y(x) & \overline{y}(x) & \overline{y}(x/q) \\
P_m(x) & \overline{P_m}(x) & \overline{P_m}(x/q) \\
Y(x)Q_n(x) &\overline{Y}(x)\overline{Q_n}(x) &\overline{Y}(x/q)\overline{Q_n}(x/q)
\end{array}
=0.
\end{equation}

\begin{theo}
The explicit form of the  equations $L_2$ and $L_3$ are obtained as
\begin{eqnarray}
& &L_2: g(a_4x)_1y(x)-(a_1x)_1y(qx)+c_1(xf)_1\overline y(x)=0 \label{qp6l2} ,\\
& &L_3 : c_2\bigg(\frac{x \overline f}q\bigg)_1y(x)+g(a_2x)_1\overline y(x)-q^{N+1}\bigg(\frac{a_3}q x\bigg)_1\overline y\bigg(\frac x q \bigg)=0\label{qp6l3}
\end{eqnarray}
where $f$, $g$, $c_1$ and $c_2$ are constant with respect to $x$ and $\overline{f} = T(f)$.
\end{theo}
\proof\\
First, we will consider the equation $L_2$. By definition of $Y(x)$, we have
\begin{eqnarray}
\frac{Y(qx)}{Y(x)}=\frac{(a_3x, a_4x)_1}{(a_1x, a_2x)_1}, && \frac{\overline{Y}(x)}{Y(x)}=\frac{(a_4x)_1}{(a_2x)_1}\nonumber.
\end{eqnarray}
Then the coefficient of $\overline{y}(x)$ in $L_2$ is computed as follows,
\begin{eqnarray*}
\lefteqn{Y(qx)P_m(x)Q_n(qx)-Y(x)P_m(qx)Q_n(x)}\\
&&=Y(x)\bigg\{ \frac{Y(qx)}{Y(x)}P_m(x)Q_n(qx)-P_m(qx)Q_n(x)\bigg\} \\
 & &=Y(x)\bigg\{\frac{(a_3x, a_4x)_1}{(a_1x, a_2x)_1}P_m(x)Q_n(qx)-P_m(qx)Q_n(x)\bigg\} \\
  & &=\frac{Y(x)}{(a_1x, a_2x)_1}\bigg\{ (a_3x, a_4x)_1P_m(x)Q_n(qx)-(a_1x, a_2x)_1P_m(qx)Q_n(x)\bigg\} .
\end{eqnarray*}
The part in \{\quad\} of the last expression is a polynomial in $x$ of order $N+2$.  By definition for Pad\'{e} approximation, this polynomial has zero at $x=0$ of order $N+1$. So, the coefficient of $\overline{y}(x)$ is given by
\begin{equation}\label{f}
A(x):=Y(qx)P_m(x)Q_n(qx)-Y(x)P_m(qx)Q_n(x) = \frac{c_0 x^{N+1}Y(x)}{(a_1x, a_2x)_1}(xf)_1
\end{equation}
where $c_0$ and $f$ are some constant.\\
 Similarly, the coefficients of $y(qx)$ and $y(x)$ are given by
\begin{eqnarray}
&&B(x):=\overline Y(x)P_m(x)\overline Q_n(x)-Y(x)\overline P_m(x)Q_n(x) = \frac{cY(x)x^{N+1}}{(a_2x)_1}, \\
&&C(x):=\overline Y(x)P_m(qx)Q_n(x)-Y(qx)\overline P_m(x)Q_n(qx) =\frac{c'Y(x)(a_4 x)_1x^{N+1}}{(a_1x, a_2x)_1}
\end{eqnarray}
where $c$,  $c'$ are some constant. By putting $g=c'/c$, $c_1=c_0/c$, we obtain $L_2$ in the equation (\ref{qp6l2}).

Next, we will consider the equation $L_3$. By definition.
\begin{eqnarray*}
\frac{Y( x/q)}{Y(x)} =\frac{\bigg(\displaystyle{\frac{a_1} q x, \frac{a_2} q x}\bigg)_1}{\bigg(\displaystyle{\frac{a_3}q x, \frac{a_4}q x}\bigg)_1}, &&
\frac{\overline Y( x/q)}{Y(x)} =\frac{\bigg(\displaystyle{\frac{a_1} q x}\bigg)_1}{\bigg(\displaystyle{\frac{a_3}q x}\bigg)_1}.
\end{eqnarray*}
The coefficient of $y(x)$ is
\begin{align*}
\overline Y(x/q)\overline P_m(x)\overline Q_n(x/q)-\overline Y(x)\overline P_m(x/q)\overline Q_n(x) &=-\overline{A}\bigg(\frac{x}{q}\bigg)\\
 &=-\frac{\overline{c_0}Y(x)}{\bigg(\displaystyle{\frac{a_3}q x, a_2 x}\bigg)_1}\bigg(\frac{x}{q}\bigg)^{N+1}\bigg(\frac{x\overline f}{q}\bigg)_1.
\end{align*}
The coefficient of $\overline{y}(x)$ is
\begin{align*}
\overline Y(x/q)P_m(x)Q/n(x/q)-Y(x)\overline P_m(x/q)Q_n(x) &=C\bigg(\frac{x}{q}\bigg)\\
 &=\frac{c'Y(x)}{\bigg(\displaystyle{\frac{a_3}qx}\bigg)_1}\bigg(\frac{x}{q}\bigg)^{N+1}.
\end{align*}
The coefficient of $\overline{y}(x/q)$ in $L_3$ is $B(x)$. So, we get $L_3$ when we simplify coefficients where $c_2=\overline{c_0}/c$.\qquad \hfill$\Box$

\subsection{Compatibility condition}\label{subseccc}

We will consider the compatibility condition of the equations (\ref{qp6l2}) and (\ref{qp6l3}).
\begin{theo}\label{ccqp6}
The compatibility condition of the equations (\ref{qp6l2}) and (\ref{qp6l3}) is given by
\begin{eqnarray}
& &g\overline g=\frac{q^{N+1}\displaystyle{\bigg(\frac{a_1}{\overline f}, \frac{a_3}{\overline f}\bigg)_1}}{\displaystyle{\bigg(\frac{qa_2}{\overline f}, \frac{qa_4}{\overline f}\bigg)_1}}\label{qp6g},\\
& &f\overline f=\frac{a_1a_3\displaystyle{\bigg(\frac{a_4}{a_1q^m}g, \frac{a_2}{a_3q^n}g\bigg)_1}}{\displaystyle{\bigg(g, \frac{g}{q^{N+1}}\bigg)_1}}\label{qp6f}.
\end{eqnarray}
This is a $q$-Painlev\'{e}VI equation\cite{jimbosakai}.
\end{theo}
\proof\\
First, we will prove the equation (\ref{qp6g}). Translate $L_2$ by $T$, we have
\begin{equation}
\overline{L_2} : \overline g(qa_4 x)_1\overline y(x)-(a_1x)_1\overline y(qx)+\overline{c_1}(x\overline f)_1\overline{\overline y}(x)=0.\nonumber
\end{equation}
We put $x=1/\overline f$ in this equation, then we have
\begin{equation}
\overline g\bigg(\frac{qa_4}{\overline f} \bigg)_1\overline y\bigg(\frac{1}{\overline f}\bigg)-\bigg(\frac{a_1}{\overline f}\bigg)_1\overline y\bigg(\frac{q}{\overline f}\bigg)=0.
\end{equation}

On the other hand, putting $x=q/\overline f$ in $L_3$, we have
\begin{equation}
g\bigg(\frac{qa_2}{\overline f}\bigg)_1\overline y\bigg(\frac{q}{\overline f}\bigg)-q^{N+1}\bigg(\frac{a_3}{\overline f}\bigg)_1\overline y\bigg(\frac {1}{\overline f} \bigg)=0.
\end{equation}
Hence,
\begin{equation}
g\overline g=\frac{q^{N+1}\displaystyle{\bigg(\frac{a_1}{\overline f}, \frac{a_3}{\overline f}\bigg)_1}}{\displaystyle{\bigg(\frac{qa_2}{\overline f}, \frac{qa_4}{\overline f}\bigg)_1}}.
\end{equation}

Then, we will prove the equation (\ref{qp6f}). $P_m(x)$ is a solution of $L_2$ and $L_3$,  so we substitute for $P(x)=k_0+ \cdots + k x^m$ in the $L_2$ and $L_3$ and check the highest order 
\begin{eqnarray*}
(-ga_4+a_1q^m)k-c_1f\overline k=0, \\
-\frac{c_2\overline f}{q}k+(-ga_2+q^na_3)\overline k=0
\end{eqnarray*}
hence, 
\begin{equation}
c_1c_2f\overline f=q^{N+1}a_1a_3\bigg(\frac{a_4}{a_1q^m}g, \frac{a_2}{a_3q^n}g\bigg)_1\label{qp6f1}.
\end{equation}
By checking the lowest order similarly, we have
\begin{equation}
c_1c_2=q^{N+1}\bigg(g, \frac{g}{q^{N+1}}\bigg)_1.\label{qp6f2}
\end{equation}
From (\ref{qp6f1}) and (\ref{qp6f2}), we get
\begin{equation}
f\overline f=\frac{a_1a_3\displaystyle{\bigg(\frac{a_4}{a_1q^m}g, \frac{a_2}{a_3q^2}g\bigg)_1}}{\displaystyle{\bigg(g, \frac{g}{q^{N+1}}\bigg)_1}}.
\end{equation}
\hfill $\Box$

\section{The case of $E^{(1)}_6$}\label{sece6}

In this section, we will derive the Lax pair for $E^{(1)}_6$ $q$-Painlev\'{e} equation (Theorem \ref{subseccdee6}), the $q$-Painlev\'{e} equation (Theorem \ref{subseccce6}) and its special solutions in a determinant formula (Theorem \ref{theoremfandg}).

\subsection{Construction of the difference equation}\label{subseccdee6}
Let $a_1, \cdots, a_4 \in \mathbb{C}\backslash \{ 0 \}$ be complex parameters. Define a function $Y(x)$ as
\begin{equation}
Y(x)=\frac{(a_1x,a_2x,a_3,a_4)_{\infty}}{(a_1,a_2,a_3x,a_4x)_{\infty}}.
\end{equation}
When
\begin{equation}\label{xi}
x_i:=q^i,
\end{equation}
we get
\begin{equation}
y_i:=Y(x_i)=\frac{(a_3, a_4)_i}{(a_1, a_2)_i}\label{yi}.
\end{equation}
Interpolate $Y(x)$ by rational function
\begin{equation} \label{padeproblem}
Y(x)=\frac{P_m(x)}{Q_n(x)}  \mbox{ ( $x=q^i$, $i=0,1,\cdots, N$ )}
\end{equation}
where $P_m(x)$ and $Q_n(x)$ are the same form in section \ref{secqp6}.
We define the transformation $T$ as
\begin{equation}\label{transformation}
T: m\mapsto m-1, \quad a_2\mapsto qa_2.
\end{equation}
We will consider the equations $L_2$, $L_3$ in the equations  (\ref{l2}) and (\ref{l3}).
\begin{theo}
The explicit form of the equations $L_2$ and $L_3$ are given as
\begin{eqnarray}
& &L_2: \bigg(\frac x g\bigg)_1y(x)-(a_1x)_1y(qx)+c_1x(xf)_1\overline y(x)=0 \label{e6l2},\\
& &L_3: c_2x\bigg(\frac{x \overline f}q\bigg)_1y(x)+\bigg(a_2x, \frac x{q^{N}}, \frac x{qg}\bigg)_1\overline y(x)-\bigg(\frac{a_3}q x, \frac{a_4}q x,x\bigg)_1\overline y\bigg(\frac x q \bigg)=0 .\label{e6l3}
\end{eqnarray}
\end{theo}
\proof\\
First, we will consider the equation $L_2$. By definition, 
\begin{eqnarray*}
\frac{Y(qx)}{Y(x)}=\frac{(a_3x, a_4x)_1}{(a_1x, a_2x)_1}, && \frac{\overline{Y}(x)}{Y(x)}=\frac{(a_2)_1}{(a_2x)_1}.
\end{eqnarray*}
The coefficient of $\overline y(x)$ is same form in $D^{(1)}_5$. So,
\begin{eqnarray*}
& &Y(qx)P_m(x)Q_n(qx)-Y(x)P_m(qx)Q_n(x) =\frac{Y(x)K(x)}{(a_1x, a_2x)_1},
\\& &K(x)=(a_3x, a_4x)_1P_m(x)Q_n(qx)-(a_1x, a_2x)_1P_m(qx)Q_n(x)
\end{eqnarray*}
where $K(x)$ is a polynomial of $x$ of order $N+2$.

By definition for Pad\'{e} interpolation, The polynomial $K(x)$ have simple zeroes at  $x=0$ and $x=q^i\quad (i=0, \cdots, N-1)$. So, the coefficient of $\overline y(x)$ is given by
\begin{equation}\label{fa}
A'(x):=Y(qx)P_m(x)Q_n(qx)-Y(x)P_m(qx)Q_n(x) = \frac{c_0 xY(x)}{(a_1x, a_2x)_1}\prod_{i=0}^{N-1}\bigg(\frac{x}{q^i}\bigg)_1(xf)_1
\end{equation}
where $c_0$ and $f$ are some constants. Similarly, the coefficients of $y(qx)$ and $y(x)$ are given by
\begin{eqnarray}
& & \!\!\!\!\!B'(x):=\overline Y(x)P_m(x)\overline Q_n(x)-Y(x)\overline P_m(x)Q_n(x)=\frac{cY(x)}{(a_2x)_1} \prod_{i=0}^{N-1}\bigg(\frac x {q^i}\bigg)_1,\nonumber \\
& &\!\!\!\!\!C'(x):=\overline Y(x)P_m(qx)Q_n(x)-Y(qx)\overline P_m(x)Q_n(qx)= \frac{c'Y(x)}{(a_1x, a_2x)_1}\prod_{i=0}^{N-1}\bigg(\frac x {q^i}\bigg)_1\bigg( \frac x g \bigg)_1 \label{g}
\end{eqnarray}
where $c$, $c'$ and $g$ are some constants.  So, we get
\begin{equation}\label{l2c}
c'\bigg(\frac x g\bigg)_1y(x)-c(a_1x)_1y(qx)+c_0x(xf)_1\overline y(x)=0.
\end{equation}
We substitute $x=0$. Then we get $c'=c$. Finally, by putting $c_1=c_0/c$, we obtain $L_2$.

Next, we will consider the equation $L_3$. Similarly in $D^{(1)}_5$, the coefficients of $y(x)$, $\overline{y}(x)$ and $\overline{y}(x/q)$ are given by $-\overline{A'}(x/q)$, $C'(x/q)$ and $B'(x)$ respectively. So, we obtain $L_3$. \hfill $\Box$

\subsection{Compatibility condition}\label{subseccce6}
We will consider the compatibility condition of the equations (\ref{e6l2}) and (\ref{e6l3}).
\begin{theo}\label{cce6}
The compatibility condition of the equations (\ref{e6l2}) and (\ref{e6l3}) is given by
\begin{eqnarray}
& &\bigg(\frac 1{fg},\frac {1}{ f \underline g}\bigg)_1 =\frac{\bigg(\displaystyle{\frac{a_1}{f}, \frac{q}{ f}, \frac{a_3}{ f}, \frac{a_4}{ f}}\bigg)_1}{\bigg(\displaystyle{\frac{a_2}{ f}, \frac {1}{q^{N} f}}\bigg{)}_1},\label{e6g}\\
& &\frac{(fg, \overline fg)_1}{f\overline f}=\frac{q^{N-1}(a_1g, qg, a_3g, a_4g)_1}{a_2\bigg(\displaystyle{ a_1 q^m g, \frac{a_3 a_4 q^n}{a_2}g}\bigg)_1}\label{e6f}.
\end{eqnarray}
This is a $q$-Painlev\'{e} equation type $E^{(1)}_6$.
\end{theo}
\proof\\
First, we will prove the equation (\ref{e6g}). We put $x=1/f$ in $L_2$.
\begin{equation}
\bigg(\frac 1 { f g}\bigg)_1 y\bigg(\frac 1 f\bigg)-\bigg(\frac{a_1}{ f}\bigg)_1 y\bigg(\frac q { f}\bigg)=0.
\end{equation}
On the other hand, Translate $L_3$ by $T^{-1}$, we have
\begin{equation}
\underline{L_3} : \underline{c_2}x\bigg(\frac{xf}{q}\bigg)_1\underline{y}(x)+\bigg(\frac{a_2}{q}x, \frac{x}{q^{N+1}}, \frac{x}{q\underline g}\bigg)_1y(x)-\bigg(\frac{a_3}{q}x, \frac{a_4}{q}x, x\bigg)_1y\bigg(\frac{x}{q}\bigg)=0.\nonumber
\end{equation}
Putting $x=q/f$ in $\underline{L_3}$, we have
\begin{equation}
\bigg(\frac{a_2}{f}, \frac{1}{q^{N}f}, \frac{1}{f\underline g}\bigg)_1y\bigg(\frac{q}{f}\bigg)-\bigg(\frac{a_3}{f}, \frac{a_4}{f}, \frac{q}{f}\bigg)_1y\bigg(\frac{1}{f}\bigg)=0.
\end{equation}
Hence,
\begin{equation}
\bigg(\frac 1{fg},\frac {1}{ f \underline g}\bigg)_1 =\frac{\bigg(\displaystyle{\frac{a_1}{f}, \frac{q}{ f}, \frac{a_3}{ f}, \frac{a_4}{ f}}\bigg)_1}{\bigg(\displaystyle{\frac{a_2}{ f}, \frac {1}{q^{N} f}}\bigg{)}_1}.
\end{equation}
Then, we prove the equation (\ref{e6f}). We put $x=g$ in $L_2$, we have
\begin{equation}
-(a_1g)y(qg)+c_1g(fg)_1\overline y(g)=0.\nonumber
\end{equation}
On the other hand, putting $x=qg$ in $L_3$, we have
\begin{equation}
c_2qg(\overline fg)_1y(qg)-(a_3 g, a_4g, q g)_1\overline y(g)=0.\nonumber
\end{equation}
Hence,
\begin{equation}
c_1c_2qg^2(fg, \overline fg)_1=(a_1g, a_3g, a_4g, qg)_1\label{e6c1c2}.
\end{equation}
We substitute for $P(x)=\cdots+kx^m$ in the $L_2$ and $L_3$ and check the highest order.
\begin{eqnarray}
\bigg(a_1q^m-\frac 1 g\bigg)k-c_1f \overline k=0, \label{e6c1}\\
\frac{c_2\overline f}qk+\bigg(\frac{a_2}{q^{N+1}g}-\frac{a_3a_4}{q^{m+1}}\bigg)\overline k=0.\label{e6c2}
\end{eqnarray}
Then, we have
\begin{equation}
c_1c_2=\frac{a_2\bigg(\displaystyle{a_1q^mg, \frac{a_3a_4q^n}{a_2}g}\bigg)_1}{q^N g^2f\overline f},\label{e6c1c22}
\end{equation}
From (\ref{e6c1c2}) and (\ref{e6c1c22}), we get
\begin{equation}
\frac{(fg, \overline fg)_1}{f\overline f}=\frac{q^{N-1}(a_1g, a_3g, a_4g, qg)_1}{a_2\bigg(\displaystyle{ a_1 q^m g, \frac{a_3 a_4 q^n}{a_2}g}\bigg)_1}.
\end{equation}
\hfill $\Box$
\subsection{Explicit form for $f$ and $g$}\label{subsecfandg}
We will derive the concrete form of $P_m$ and $Q_n$ (Lemma \ref{pandg}). Then, by using these formulas, we will derive the explicit form of $f$ and $g$ (Theorem \ref{theoremfandg}).

First, we consider the Pad\'{e} problem (\ref{padeproblem}) for general $ \{x_i\}$ and $\{y_i\}$. We will prove following theorem.

\begin{theo}
The polynomials $P_m(x)$ and $Q_n(x)$ are given as
\begin{eqnarray}
& &P_m(x)=F(x) \det\bigg(\sum_{s=0}^N x_s^{i+j}\frac{u_s}{x-x_s}\bigg)_{i, j=0}^{n}, \label{padep}\\
& &Q_n(x)=\det\bigg(\sum_{s=0}^N x_s^{i+j}u_s (x-x_s)\bigg)_{i, j=0}^{n-1}
\end{eqnarray}
where
$u_s=y_s/F'(x_s)$, $F(x)=\prod_{i=0}^N(x-x_i)$.
\end{theo}
\proof\\
In this proof, we use the notation $[k]=\{0, 1, \cdots, k\}$ for $k \in \mathbb{Z}_{\geq 0}$ and $\Delta_{[k]}=\prod_{\substack{\alpha, \beta \in [k] \\ \alpha < \beta}}(x_{\alpha}-x_{\beta})$.

The determinant in the equation (\ref{padep}) is evaluated as
\begin{align*}
\det\bigg(\sum_{s=0}^N x_s^{i+j}\frac{u_s}{x-x_s}\bigg)_{i, j=0}^{n} &= \bigg|(x^i _s)^{i\in [n]}_{s\in [N]}\mathrm{diag}\bigg(\frac{u_s}{x-x_s}\bigg)^{N}_{s=0} (x^j_s)^{s\in [N]}_{j\in [n]}\bigg| \\
&=\sum_{\substack{I\subset[N]\\ |I|=n+1}}\prod_{s\in I}\frac{u_s}{x-x_s}\bigg|(x_s^i)^{i\in [n]}_{s\in I}\bigg| \bigg|(x^j_s)^{s\in I}_{j\in [n]}\bigg|\\
&= \sum_{\substack{I\subset [N]\\ |I|=n+1}}\prod_{s\in I}\frac{u_s}{x-x_s} \Delta_{I}^2.
\end{align*}
Similarly,
\begin{equation}
\det\bigg(\sum_{s=0}^N x_s^{i+j}u_s (x-x_s)\bigg)_{i, j=0}^{n-1}= \sum_{\substack{I\subset [N]\\ |I|=n}}\Delta_{I}^2\prod_{s\in I}u_s (x-x_s).\nonumber
\end{equation}
Let
\begin{eqnarray*}
R(x)= \sum_{\substack{I\subset [N]\\ |I|=n+1}}\Delta_{I}^2\prod_{s\in I}\frac{u_s}{x-x_s}, &&W(x)=\sum_{\substack{I\subset [N]\\ |I|=n}}\Delta_{I}^2\prod_{s\in I}u_s (x-x_s) 
\end{eqnarray*}
and we will prove
\begin{equation}\label{padegeneral}
\frac{F(x)R(x)}{W(x)}\bigg|_{x=x_i}=y_i.
\end{equation}
We substitute $x=x_i (i\in [N])$ in $F(x)R(x)$ and put  $I=I'\cup \{ i \}$, then
\begin{align*}
F(x_i)R(x_i) &=\sum_{\substack{I\subset [N]\\ |I|=n+1}}\Delta_{I}^2 \prod_{s\in I}u_s\prod_{s\not\in I}(x_i-x_s) \\
&=\sum_{\substack{I' \cup\{ i \}\subset [N] \\ |I'\cup\{ i \}|=n+1}}\Delta_{I'\cup \{i\}}^2\prod_{s\in I'\cup\{i\}}u_s \prod_{s\not\in I'\cup\{i\}}(x_i-x_s)\\
&=\sum_{\substack{I'\cup\{i \} \subset [N]\\ |I'|=n}} \Delta_{I'\cup \{ i \}}^2u_i\prod_{s\in I'}u_s\prod_{\substack{s\not\in I'\\ s\neq i}}(x_i-x_s) \\
&=u_i \sum_{\substack{I'\cup\{i \} \subset [N]\\ |I'|=n}} \Delta_{I'}^2\prod_{s\in I'}(x_i-x_s)^2 \prod_{s\in I'}u_s\prod_{\substack{s\not\in I'\\s\neq i}}(x_i-x_s)\allowdisplaybreaks\\
&=u_iF'(x_i)\sum_{\substack{I'\cup\{i \} \subset [N]\\ |I'|=n}}\Delta_{I'}^2\prod_{s\in I'}(x_i-x_s)\prod_{s\in I'}u_s, \\
W(x_i) &= \sum_{\substack{I\subset [N]\\ |I|=n}}\Delta_{I}^2\prod_{s\in I}u_s (x_i-u_s) \\
&=\sum_{\substack{I'\subset [N]\\ |I'|=n}}\Delta_{I'}^2\prod_{s\in I'}u_s(x_i-x_s)\quad(\text{if}\quad s=i, \; \prod_{s\in I}u_s(x_i-x_s)=0).
\end{align*}
So,
\begin{equation}
\frac{F(x_i)R(x_i)}{W(x_i)}=u_iF'(x_i)=y_i.\nonumber
\end{equation}
Hence, (\ref{padegeneral}) is proved. \hfill $\Box$

\begin{lemm}\label{pandg}
By specializing $x_i$ and $y_i$ as the equations (\ref{xi}) and (\ref{yi}) respectively, we have
\begin{eqnarray}
& &P_m(x)=F(x)\det\bigg( \frac{(q^{N+1})_{\infty}}{(q)_{\infty}}\sum_{s=0}^N \frac{(a_3, a_4, q^{-N})_s}{(a_1, a_2, q)_s}\frac{q^{s(i+j+1)}}{x-q^s}\bigg)_{i, j=0}^n \label{pdet},\\
& &Q_n(x)=\det\bigg(\frac{(q^{N+1})_{\infty}}{(q)_{\infty}} \sum_{s=0}^N \frac{(a_3, a_4, q^{-N})_s}{(a_1, a_2, q)_s}q^{s(i+j+1)}(x-q^s)\bigg)_{i, j=0}^{n-1}\label{qdet}.
\end{eqnarray}
\end{lemm}
\proof\\
First, we calculate $F'(x_s)$.
\begin{align*}
F'(x_s)&=(x_s-x_0)\cdots(x_s-x_{s-1})(x_s-x_{s+1})\cdots(x_s-x_N)\\
&=(-1)^sq^{1/2(s-1)s}(q)_sq^{s(N-s)}(q)_{N-s}.
\end{align*}
Here,
\begin{align*}
(q)_{N-s}&=\frac{(q)_{\infty}}{(q^{N-s+1})_{\infty}}\\
&=\frac{(q)_{\infty}}{(q^{N-s+1})_{s}(q^{N+1})_{\infty}}
\end{align*}
Furthermore, 
\begin{equation}\label{q-Nrelation}
(q^{-N})_s=(-1)^sq^{(s-1)s/2-Ns}(q^{N-s+1})_s.
\end{equation}
So,
\begin{equation}\label{(q)relation}
(q)_{N-s}=(-1)^sq^{s(s-1)/2-Ns}\frac{(q)_{\infty}}{(q^{-N})_s(q^{N+1})_{\infty}}.
\end{equation}
By using this relation (\ref{(q)relation}), $F'(x)$ is given by
\begin{equation}\label{F'}
F'(x_s)=q^{-s}\frac{(q)_s(q)_{\infty}}{(q^{-N})_s (q^{N+1})_{\infty}}.
\end{equation}
By using the equation (\ref{F'}), the equation (\ref{pdet}) is evaluated as
\begin{align*}
P_m(x)&=F(x)\det\bigg(\sum_{s=0}^N x_s^{i+j}\frac{u_s}{x-x_s}\bigg)_{i, j=0}^{n} \\
&=F(x)\det\bigg(\frac{(q^{N+1})_{\infty}}{(q)_{\infty}} \sum_{s=0}^N \frac{(a_3, a_4, q^{-N})_s}{(a_1, a_2, q)_s}\frac{q^{s(i+j+1)}}{x-q^s}\bigg)_{i, j=0}^n.
\end{align*}

$Q_n(x)$ is similar.\hfill$\Box$

By using the Lemma \ref{pandg},  we get the following lemma.
\begin{lemm}\label{qhge}
The values of polynomials $P_m(x)$ and $Q_n(x)$ at special points are given as follows.
\begin{eqnarray*}
& &P_m\bigg(\frac{1}{a_1}\bigg) =\frac{(a_1)_{N+1}}{a_1^m}\det \bigg(\frac{(q^{N+1})_{\infty}}{(a_1)_1(q)_{\infty}}{_{3}\varphi_2}\big(\substack{\displaystyle{a_3, a_4, q^{-N}} \\ \displaystyle{qa_1, a_2}} ;q^{i+j+1}\big)\bigg)_{i,j=0}^{n},\\
& &P_m\bigg(\frac{1}{a_2}\bigg) =\frac{(a_2)_{N+1}}{a_2^m}\det \bigg(\frac{(q^{N+1})_{\infty}}{(a_2)_1(q)_{\infty}}{_{3}\varphi_2}\big(\substack{\displaystyle{a_3, a_4, q^{-N}} \\ \displaystyle{  a_1, qa_2}} ;q^{i+j+1}\big)\bigg)_{i,j=0}^{n},\\
& &P_m\bigg(\frac{q}{a_3}\bigg) =\bigg(\frac{q}
{a_3}\bigg)^m\bigg(\frac{a_3}{q}\bigg)_{N+1}\det\bigg(\frac{(q^{N+1})_{\infty}}{(a_3/q)_1 (q)_{\infty}}{_{3}\varphi_2}\big(\substack{\displaystyle{a_3/q, a_4, q^{-N}} \\\displaystyle{ a_1, a_2}
 };q^{i+j+1}\big)\bigg)_{i,j=0}^{n},\\
& &P_m\bigg(\frac{q}{a_4}\bigg) =\bigg(\frac{q}{a_4}\bigg)^m\bigg(\frac{a_4}{q}\bigg)_{N+1}\det\bigg(\frac{(q^{N+1})_{\infty}}{(a_4/q)_1(q)_{\infty}}{_{3}\varphi_2}\big(\substack{\displaystyle{a_3, a_4/q, q^{-N}}\\ \displaystyle{ a_1, a_2}} ;q^{i+j+1}\big)\bigg)_{i,j=0}^{n},\allowdisplaybreaks\\
& &Q_n\bigg(\frac{q}{a_1}\bigg) =\bigg(\frac{q}{a_1}\bigg)^{n}\det\bigg( \bigg(\frac{a_1}{q}\bigg)_1\frac{(q^{N+1})_{\infty}}{(q)_{\infty}}{_{3}\varphi_2}\big(\substack{\displaystyle{a_3, a_4, q^{-N}} \\ \displaystyle{ a_1/q, a_2}} ;q^{i+j+1}\big)\bigg)_{i,j=0}^{n-1},\allowdisplaybreaks\\
& &Q_n\bigg(\frac{q}{a_2}\bigg) =\bigg(\frac{q}{a_2}\bigg)^{n}\det\bigg(\bigg(\frac{a_2}{q}\bigg)_1\frac{(q^{N+1})_{\infty}}{(q)_{\infty}}{_{3}\varphi_2}\big(\substack{\displaystyle{a_3, a_4, q^{-N} }\\  \displaystyle{a_1, a_2/q}} ;q^{i+j+1}\big)\bigg)_{i,j=0}^{n-1},\\
& &Q_n\bigg(\frac{1}{a_3}\bigg) =\bigg(\frac{1}{a_3}\bigg)^{n}\det\bigg( (a_3)_1 \frac{(q^{N+1})_{\infty}}{(q)_{\infty}} {_{3}\varphi_2}\big(\substack{\displaystyle{qa_3, a_4, q^{-N}}\\ \displaystyle{a_1, a_2}} ;q^{i+j+1}\big)\bigg)_{i,j=0}^{n-1},\\
& &Q_n\bigg(\frac{1}{a_4}\bigg) =\bigg(\frac{1}{a_4}\bigg)^n \det\bigg( (a_4)_1 \frac{(q^{N+1})_{\infty}}{(q)_{\infty}}{_{3}\varphi_2}\big(\substack{\displaystyle{a_3, qa_4, q^{-N}} \\ \displaystyle{ a_1, a_2}} ;q^{i+j+1}\big)\bigg)_{i,j=0}^{n-1}
\end{eqnarray*}
where
\begin{equation}\label{qhgs}
_3\varphi_2 \big(\substack{\displaystyle{a_1, a_2, a_3}\\{\displaystyle{b_1, b_2}}}; x \big)=\sum_{s=0}^{\infty}\frac{(a_1, a_2, a_3)_s}{(b_1, b_2, q)_s }x^s.
\end{equation}
$_3\varphi_2 $ is a $q$-hypergeometric series.
\end{lemm}
\proof\\
We prove $P_m(1/a_1)$. Using the equation (\ref{pdet}),
\begin{align*}
P_m\bigg(\frac{1}{a_1}\bigg)
&=F\bigg(\frac{1}{a_1}\bigg)\det\bigg( \frac{(q^{N+1})_{\infty}}{(q)_{\infty}}\sum_{s=0}^N \frac{(a_3, a_4, q^{-N})_s}{(a_1, a_2, q)_s}\frac{q^{s(i+j+1)}}{1/a_1-q^s}\bigg)_{i, j=0}^n \\
&=\prod_{i=0}^{N}\bigg(\frac{1}{a_1}-q^i\bigg)\det\bigg( \frac{(q^{N+1})_{\infty}}{(q)_{\infty}}\sum_{s=0}^N \frac{(a_3, a_4, q^{-N})_s}{(a_1, a_2, q)_s}\frac{q^{s(i+j+1)}}{\displaystyle{\frac{1-a_1q^s}{a_1}}}\bigg)_{i, j=0}^n \\
&=\bigg(\frac{1}{a_1}\bigg)^{N+1}\prod_{i=0}^{N}(1-q^ia_1)\det\bigg(\frac{a_1}{1-a_1} \frac{(q^{N+1})_{\infty}}{(q)_{\infty}} \sum_{s=0}^N \frac{(a_3, a_4, q^{-N})_s}{(q a_1, a_2, q)_s}q^{s(i+j+1)}\bigg)_{i, j=0}^n \\
&=\frac{(a_1)_{N+1}}{a_1^m}\det \bigg(\frac{(q^{N+1})_{\infty}}{(a_1)_1(q)_{\infty}}{_{3}\varphi_2}\big(\substack{\displaystyle{a_3, a_4, q^{-N}} \\ \displaystyle{qa_1, a_2}} ;q^{i+j+1}\big)\bigg)_{i,j=0}^{n}.
\end{align*} 
Others are similar.\hfill $\Box$

By using the above Lemma \ref{qhge}, we will get the following theorem.

\begin{theo}\label{theoremfandg}
The explicit form of $f$ and $g$ are
\begin{align}
\!\!\!\!\!\frac{\bigg(\displaystyle{\frac{f}{a_1}}\bigg)_1}{\bigg(\displaystyle{\frac{f}{a_2}}\bigg)_1}
&=A\frac{\det\bigg({_3\varphi_2}\big(\substack{\displaystyle{a_3, a_4, q^{-N}}\\\displaystyle{q a_1, a_2}}; q^{i+j+1}\big)\bigg)_{i, j=0}^n\det\bigg({_3\varphi_2}\big(\substack{\displaystyle{ a_3, a_4, q^{-N}}\\ \displaystyle{a_1/q, a_2}}; q^{i+j+1}\big)\bigg)_{i, j=0}^{n-1}}{\det\bigg({_3\varphi_2}\big(\substack{\displaystyle{a_3, a_4, q^{-N}}\\\displaystyle{a_1, qa_2}}; q^{i+j+1}\big)\bigg)_{i, j=0}^n\det\bigg({_3\varphi_2}\big(\substack{\displaystyle{a_3,  a_4, q^{-N}}\\ \displaystyle{a_1, a_2/q}}; q^{i+j+1}\big)\bigg)_{i, j=0}^{n-1}}\label{fc},\\
\!\!\!\!\!\frac{(ga_3)_1}{(ga_4)_1}
&=B\frac{\det\bigg({_3\varphi_2}\big(\substack{\displaystyle{a_3/q, a_4, q^{-N}}\\\displaystyle{ a_1, a_2}}; q^{i+j+1}\big)\bigg)_{i, j=0}^n\det\bigg({_3\varphi_2}\big(\substack{\displaystyle{ qa_3, a_4, q^{-N}}\\ \displaystyle{a_1, a_2}}; q^{i+j+1}\big)\bigg)_{i, j=0}^{n-1}}{\det\bigg({_3\varphi_2}\big(\substack{\displaystyle{a_3, a_4/q, q^{-N}}\\\displaystyle{a_1, a_2}}; q^{i+j+1}\big)\bigg)_{i, j=0}^n\det\bigg({_3\varphi_2}\big(\substack{\displaystyle{a_3, qa_4, q^{-N}}\\ \displaystyle{a_1, a_2, q}}; q^{i+j+1}\big)\bigg)_{i, j=0}^{n-1}}\label{gc}
\end{align}
where
\begin{eqnarray*}
& &A=\frac{a_1}{a_2}\frac{\bigg(\displaystyle{\frac{a_3}{a_1}, \frac{a_4}{a_1}, \overbrace{a_2, \cdots, a_2}^{n+1}, \overbrace{\displaystyle{\frac{a_1}{q}, \cdots, \frac{a_1}{q}}}^{n}, q^Na_1}\bigg)_1}{\bigg(\displaystyle{\frac{a_3}{a_2}, \frac{a_4}{a_2}, \underbrace{a_1, \cdots, a_1}_{n+1}, \underbrace{\frac{a_2}{q}, \cdots, \frac{a_2}{q}}_{n}}, q^Na_2\bigg)_1},\allowdisplaybreaks \\
& &B=\frac{a_4}{a_3}\frac{\bigg(\displaystyle{\frac{a_1}{a_3} }, \overbrace{\displaystyle{\frac{a_4}{q}, \cdots, \frac{a_4}{q}}}^{n+1}, \overbrace{a_3, \cdots, a_3}^{n},\displaystyle{\frac{a_3}{q}}\bigg)_1}{\bigg(\displaystyle{\frac{a_1}{a_4}}, \underbrace{\displaystyle{\frac{a_3}{q}, \cdots, \frac{a_3}{q}}}_{n+1}, \underbrace{a_4, \cdots, a_4}_{n},\displaystyle{\frac{a_4}{q}}\bigg)_1}.
\end{eqnarray*}
\end{theo}
\proof\\
First, we prove the equation (\ref{fc}). We substitute $x=1/a_1, 1/a_2$ for the equation (\ref{fa}), we have
\begin{eqnarray*}
& &\bigg(\frac{a_3}{a_1}, \frac{a_4}{a_1}\bigg)_1P_m\bigg(\frac{1}{a_1}\bigg)Q_n\bigg(\frac{q}{a_1}\bigg)=\prod_{i=0}^{N-1}\bigg(\frac{1}{q^ia_1}\bigg)_1\frac{c_0}{a_1}\bigg(\frac{f}{a_1}\bigg)_1,\\
& &\bigg( \frac{a_3}{a_2}, \frac{a_4}{a_2}\bigg)_1P_m\bigg(\frac{1}{a_2}\bigg)Q_n\bigg(\frac{q}{a_2}\bigg)=\prod_{i=0}^{N-1}\bigg(\frac{1}{q^ia_2}\bigg)_1\displaystyle{\frac{c_0}{a_2}}\bigg(\frac{f}{a_2}\bigg)_1
\end{eqnarray*}
respectively. Taking a ratio of these two equations, we have
\begin{equation}
\frac{\bigg(\displaystyle{\frac{f}{a_1}}\bigg)_1}{\bigg(\displaystyle{\frac{f}{a_2}}\bigg)_1}=\bigg(\frac{a_1}{a_2}\bigg)^{N+1}\frac{\bigg(\displaystyle{\frac{a_3}{a_1}, \frac{a_4}{a_1}}\bigg)_1}{\bigg(\displaystyle{\frac{a_3}{a_2}, \frac{a_4}{a_2}}\bigg)_1}\prod_{i=0}^{N-1}\frac{(q^ia_2)_1}{(q^ia_1)_1}\frac{P_m\bigg(\displaystyle{\frac{1}{a_1}}\bigg)Q_n\bigg(\displaystyle{\frac{q}{a_1}}\bigg)}{P_m\bigg(\displaystyle{\frac{1}{a_2}}\bigg)Q_n\bigg(\displaystyle{\frac{q}{a_2}}\bigg)}.\nonumber
\end{equation}
Then, by using the Lemma \ref{qhge}, we obtain the equation (\ref{fc}).

The proof of the equation (\ref{gc}) is similar, where we substitute $x=1/a_3, 1/a_4$ for the equation (\ref{g}). \hfill $\Box$
  
\appendix

\section{Relation to the QRT system}\label{qrt}
Painlev\'{e} equation is a non-autonomaization of the QRT system\cite{RGH}. In this appendix, Using a pencil (i.e. a 1-parameter family ) on $\mathbb{P}^1\times \mathbb{P}^1$ of order $(2, 2)$,  we will derive the autonomaization of the equations (\ref{qp6g}), (\ref{qp6f}) for $q$-Painlev\'{e} VI equation and the equations (\ref{e6g}), (\ref{e6f}) for $q$-Painlev\'{e } equation of type $E^{(1)}_6$ from the QRT point of view.

\subsection{The case of $q$-Painlev\'{e}VI equation}\label{subsecqp6qrt}
\begin{figure}[h]
 \begin{center}
    \includegraphics[width=4cm]{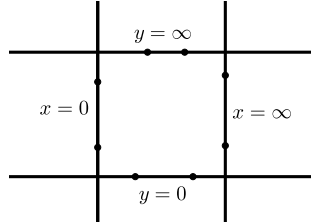}
   \caption{Eight points for $q$-Painlev\'{e}VI}\label{qp6fig}
  \end{center}
\end{figure}

We consider a polynomial of order $(2, 2)$ passing the eight points at 
\begin{equation}\label{eightpointsqp6}
(a_1, 0), (a_2, 0), (0, a_3), (0, a_4), (a_5, \infty), (a_6, \infty), (\infty, a_7), (\infty, a_8)
\end{equation}
(Figure \ref{qp6fig}). Then, the following lemma is obtained.

\begin{lemm}\label{qrtqp6lemm}
When the eight parameters $a_1$, $\cdots$, $a_8$ satisfy the condition
\begin{equation}\label{conditionqp6}
\frac{a_1a_2a_7a_8}{a_3a_4a_5a_6}=1,
\end{equation}
the polynomial of order $(2, 2)$ passing the eight points (\ref{eightpointsqp6}) forms a 1-parameter family.
\end{lemm}
\proof\\
Let $P(x)$ be a polynomial of order $(2, 2)$ passing through the eight points (\ref{eightpointsqp6}).
First, we consider the conditions that $P(x)$ passes the points at $(a_1, 0)$, $(a_2, 0)$, $(0, a_3)$, $(0, a_4)$, $(a_5, \infty)$, $(a_6, \infty)$. Then, $P(x)$ is determined as
\begin{multline}\label{qp6qrtsubstitute}
x^2-(a_1+a_2)x+a_1a_2-\frac{a_1a_2}{a_3a_4}(a_3+a_4)y+\frac{a_1a_2}{a_3a_4}y^2-\frac{a_1a_2}{a_3a_4a_5a_6}(a_5+a_6)xy^2\\+ \frac{a_1a_2}{a_3a_4a_5a_6}x^2y^2+a x^2y+\lambda xy=0
\end{multline}
up to over all constant, where $a$ and $\lambda$ are parameters. We substitute $x=1/u$ for the equation (\ref{qp6qrtsubstitute}) and put $u=0$. Then, 
\begin{equation}\label{qp6qrteqation}
\frac{a_1a_2}{a_3a_4a_5a_6}y^2+ay+1=0.
\end{equation}
By definition of the polynomial $P(x)$, the solutions of the equation (\ref{qp6qrteqation}) are $a_7$ and $a_8$. So, 
\begin{equation}
\frac{a_1a_2}{a_3a_4a_5a_6}y^2+ay+1=\frac{a_1a_2}{a_3a_4a_5a_6}(y-a_7)(y-a_8).
\end{equation}
By comparing coefficients of the equation above, we get the condition (\ref{conditionqp6}) and the relation
\begin{equation}
a=-\frac{a_1a_2(a_7+a_8)}{a_3a_4a_5a_6}.
\end{equation}
As a result, $P(x)$ is given by
\begin{equation}\label{qp6qrtpencil}
\lambda xy +F(x, y)=0
\end{equation}
where $F(x, y)$ is a polynomial of order $(2, 2)$ and $\lambda$ is a parameter. \hfill $\Box$

We consider the case where the condition (\ref{conditionqp6}) is satisfied. When we give a generic initial point $(x_0, y_0)$, then the curve $C_0$ passing through it is determined uniquely. Then we get the following theorem.

\begin{theo}\label{qp6qrttheo}
Consider the intersection of the curve $C_0$ and $\{x=x_0\}$, and let $(x_0, y_0)$ and $(x_0, y_1)$ be  the intersection points. Then, we get
\begin{equation}\label{qp6qrty}
y_0 y_1 = \frac{(x_0-a_1)(x_0-a_2)}{(x_0-a_5)(x_0-a_6)}a_7a_8,
\end{equation}
Similarly, Consider the intersection of $C_0$ and $\{y=y_0\}$, and let $(x_0, y_0)$ and $(x_1, y_0)$ be the intersection points. Then,  
\begin{equation}\label{qp6qrtx}
x_0x_1 = \frac{(y_0-a_3)(y_0-a_4)}{(y_0-a_5)(y_0-a_6)}a_5a_6.
\end{equation}
\end{theo}
\proof\\
 We prove the equation (\ref{qp6qrty}). First, the curve $C_0$ is given by the equation (\ref{qp6qrtpencil}) where $\lambda=-F(x_0, y_0)/x_0y_0$. Let $F(x, y)=A(x)y^2+B(x)y+C(x)$ where $A(x)$, $B(x)$ and $C(x)$ are second degree polynomials with respect to $x$. And we put $x=x_0$ in the curve $C_0$. Then we get a following equation 
 \begin{equation}\label{qp6qrtc0}
 A(x_0)y_0y^2-\{A(x_0)y_0^2+C(x_0)\}y+C(x_0)y_0=0.
 \end{equation}
 The solutions of the equation (\ref{qp6qrtc0}) are $y_0$ and $y_1$. So, by the relation of roots and coefficients, we get
\begin{equation}
y_0y_1=\frac{C(x_0)}{A(x_0)}.
\end{equation}
Then, by the definition for the polynomial $F(x, y)$, it follows that
\begin{eqnarray*}
F(a_i, 0)=0 \quad (i=1, 2),  && F(a_j, \infty)=0 \quad (j= 5, 6).
\end{eqnarray*}
And the solutions of the equation $F(0, y)=0$ are $a_3$, $a_4$. So, by the relation of roots and coefficients, $y_0y_1=a_3a_4$ when $x_0=0$.

By the results of above and the condition (\ref{conditionqp6}), we get the equation (\ref{qp6qrty}).  The equation (\ref{qp6qrtx}) is similar.\hfill $\Box$

The equations (\ref{qp6qrty}) and (\ref{qp6qrtx}) are the same form for the $q$-Painlev\'{e}  VI equation  (\ref{qp6g}) and (\ref{qp6f}).

\subsection{The case of $E^{(1)}_6$}\label{subsece6qrt}
\begin{figure}[h]
 \begin{center}
    \includegraphics[width=4cm]{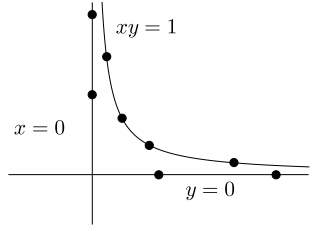}
   \caption{Eight points for $E^{(1)}_6$}\label{e6figure}
  \end{center}
\end{figure}

Similarly in appendix \ref{subsecqp6qrt},  we consider the polynomial of order $(2, 2)$ and eight points at 
\begin{equation}\label{eightpointse6}
(a_1, 0), (a_2, 0), (0, a_3), (0, a_4), (a_5, \displaystyle{\frac{1}{a_5}}), (a_6, \displaystyle{\frac{1}{a_6}}), (a_7, \displaystyle{\frac{1}{a_7}}), (a_8, \displaystyle{\frac{1}{a_8}})
\end{equation}
(Figure \ref{e6figure}). Then, the following lemma is obtained.
\begin{lemm}
When the eight parameters $a_1, a_2, \cdots, a_8$ satisfy the condition
\begin{equation}\label{conditione6}
\frac{a_3a_4a_5a_6a_7a_8}{a_1a_2}=1,
\end{equation}
the polynomial of order $(2, 2)$ passing the eight points (\ref{eightpointse6}) forms a 1-parameter family.
\end{lemm}
\proof\\
Let $P(x)$ be a polynomial of order $(2, 2)$ passing through the eight points (\ref{eightpointsqp6}). First, we consider the conditions that $P(x)$ passes the points at $(a_1, 0)$, $(a_2, 0)$, $(0, a_3)$, $(0, a_4)$. Then, $P(x)$ is determined as
\begin{equation}
x^2-(a_1+a_2)x+a_1a_2-\frac{a_1a_2}{a_3a_4}(a_3+a_4)y+\frac{a_1a_2}{a_3a_4}y^2+ax^2y^2+bx^2y+cxy^2+\lambda xy=0
\end{equation}
up to over all constant, where $a, b, c$ and $\lambda$ are parameters. We substitute $x=u$ and $y=1/u$. Then, 
\begin{equation}\label{e6qrtequation}
u^4+(b-a_1-a_2)u^3+(a_1a_2+a+\lambda)u^2+\{c-\frac{a_1a_2}{a_3a_4}(a_3+a_4)\}u+\frac{a_1a_2}{a_3a_4}=0
\end{equation}
By definition of the polynomial $P(x)$, the solutions of the equation (\ref{e6qrtequation}) are $a_5$, $a_6$, $a_7$ and $a_8$. So, 
\begin{multline}
u^4+(b-a_1-a_2)u^3+(a_1a_2+a+\lambda)u^2+\{c-\frac{a_1a_2}{a_3a_4}(a_3+a_4)\}u+\frac{a_1a_2}{a_3a_4}\\=(u-a_5)(u-a_6)(u-a_7)(u-a_8).
\end{multline}
By comparing coefficients of the equation above, we get the condition (\ref{conditione6}) and the following relations
\begin{eqnarray*}
& &a=-\lambda-a_1a_2+a_5a_6+a_5a_7+a_5a_8+a_6a_7+a_6a_8+a_7a_8,\\
& &b=a_1+a_2-(a_5+a_6+a_7+a_8),\\
& &c=\frac{a_1a_2}{a_3a_4}(a_3+a_4)-(a_5a_6a_7+a_5a_6a_8+a_5a_7a_8+a_6a_7a_8).
\end{eqnarray*}
As a result, $P(x)$ is given by
\begin{equation}\label{e6qrtpencil}
\lambda xy(1-xy)+F(x, y)=0
\end{equation}
where $F(x, y)$ is a polynomial of order $(2, 2)$ and $\lambda$ is a parameter. \hfill $\Box$

We consider the case where the condition (\ref{conditione6}) is satisfied. We give a generic initial point $(x_0, y_0)$, then the curve $C_0$ passing through it is determined uniquely. Then we get the following theorem.

\begin{theo}\label{e6qrttheo}
Consider the intersection of the curve $C_0$  and $\{x=x_0\}$, and let $(x_0, y_0)$ and $(x_0, y_1)$ be intersection points. Then, we get
\begin{equation}
\frac{(x_0y_1-1)(x_0y_0-1)}{y_0y_1} = \frac{(x_0-a_5)(x_0-a_6)(x_0-a_7)(x_0-a_8)}{(x_0-a_1)(x_0-a_2)}.\label{e6qrty}
\end{equation}
Similarly, Consider the intersection of the curve $C_0$ and $\{y=y_0\}$, and let $(x_0, y_0)$ and $(x_1, y_0)$ be intersection points. Then, 
\begin{equation}
\frac{(x_1y_0-1)(x_0y_0-1)}{x_0x_1} = \frac{(y_0a_5-1)(y_0a_6-1)(y_0a_7-1)(y_0a_8-1)}{(y_0-a_3)(y_0-a_4)}.\label{e6qrtx}
\end{equation}
\end{theo}
\proof\\
We prove the equation (\ref{e6qrty}). First, the curve $C_0$ is given by the equation (\ref{e6qrtpencil}) where $\lambda=-F(x_0, y_0)/x_0y_0(1-x_0y_0)$. Let $F(x, y)=A(x)y^2+B(x)y+C(x)$ where $A(x)$, $B(x)$ and $C(x)$ are second degree polynomials with respect to $x$. And we put $x=x_0$ in the curve $C_0$. Then, we get the following equation
\begin{multline}\label{e6qrtc0}
(A(x_0)y_0+B(x_0)x_0y_0+C(x_0)x_0)y^2-(A(x_0)y_0^2+B(x_0)x_0y_0^2+C(x_0))y\\+C(x_0)y_0(1-x_0y_0)=0.
\end{multline}
The solutions of the equation (\ref{e6qrtc0}) are $y_0$ and $y_1$. So, by the relation of roots and coefficient, we get
\begin{equation}
y_0y_1=\frac{C(x_0)y_0(1-x_0y_0)}{A(x_0)y_0+B(x_0)x_0y_0+C(x_0)x_0}.
\end{equation}
Namely,
\begin{equation}\label{e6qrtrac}
y_1=\frac{C(x_0)(1-x_0y_0)}{A(x_0)y_0+B(x_0)x_0y_0+C(x_0)x_0}.
\end{equation}
We substitute the equation (\ref{e6qrtrac}) for $(x_0y_1-1)/y_1$. Then, we get the following equation
\begin{align}
\frac{x_0y_1-1}{y_1}&=-\frac{(A(x_0)+B(x_0)x_0+C(x_0)x_0^2)y_0}{C(x_0)(1-x_0y_0)}\nonumber\\
 &=-\frac{x_0^2F(x_0, 1/x_0)y_0}{C(x_0)(1-x_0y_0)}\label{e6qrt2}.
\end{align}
The order of the numerator and the denominator of the r.h.s. of the equation (\ref{e6qrt2}) are $(4, 1)$ and $(3, 1)$ respectively. By the definition for the polynomial $F(x, y)$, it follows that
\begin{eqnarray*}
 a_i^2F(a_i, 1/a_i)=0  \quad(i=5, 6, 7, 8), && F(a_j, 0)=0 \quad (j=1, 2).
 \end{eqnarray*}
and hence, $y_0y_1=a_3a_4$ when $x_0=0$.
 
By the results of above and the condition (\ref{conditione6}), we get the equation (\ref{e6qrty}). The equation (\ref{e6qrtx}) is similar.  \hfill $\Box$

The equations (\ref{e6qrty}) and (\ref{e6qrtx}) are same form for the $q$-Painlev\'{e} equation of type $E^{(1)}_6$ (\ref{e6g}) and (\ref{e6f}). 

\section{Pad\'{e} Method and B\"{a}cklund transformations}\label{backlund}
In the main text, we constructed the $q$-Painlev\'{e} equation of type $E^{(1)}_6$ with respect to the  direction (\ref{transformation}). But, we can construct $q$-Painlev\'{e} equation by using other directions.

In this appendix, we will discuss the relation between these $q$-Painlev\'{e} equations along the various direction and B\"{a}cklund transformations.
First, we will summarize about B\"{a}cklund transformations. Then, we will discuss the relation between the $q$-Painlev\'{e} equations for various directions and B\"{a}cklund transformations.

\subsection{B\"{a}cklund transformations}\label{subsecbt}
We formulate the B\"{a}cklund transformations of the affine Weyl group of type $E^{(1)}_6$.

Let $A$ be the Cartan matrix of type $E^{(1)}_6$:
\begin{equation}
A=(a_{ij})_{i, j=0}^6=
\left(
\begin{array}{ccccccc}
2 & 0 & 0 & 0 & 0 & 0 & -1\\
0 & 2 &-1 & 0 & 0 & 0 & 0 \\  
0 & -1 &2 & -1 & 0 & 0 & 0 \\
0 & 0 & -1 &2 & -1 & 0 & -1 \\  
0 & 0 & 0 & -1 &2 & -1 & 0 \\
 0 & 0 & 0 & 0 & -1 &2 & 0 \\
-1 & 0 & 0 & -1 & 0 & 0 & 2\\
\end{array}
\right).\nonumber
\end{equation}

We define the transformations $s_i$ ($i=0, 1, 2, 3, 4, 5, 6$) and $\pi_j$ ($j=1, 2$) on the parameters $b_k$ ($k=1, 2, 3, 4, 5, 6, 7, 8$) and variables $f$ and $g$ as follows.
\begin{eqnarray}
&s_0 : (b_1, b_2, b_3, b_4, b_5, b_6, b_7, b_8, f, g)& \!\!\!\!\mapsto  (b_1, b_2, b_4, b_3, b_5, b_6, b_7, b_8, f, g),\nonumber\\
&s_1 : (b_1, b_2, b_3, b_4, b_5, b_6, b_7, b_8, f, g)&\!\!\!\!\mapsto  (b_2, b_1, b_3, b_4, b_5, b_6, b_7, b_8, f, g),\nonumber\allowdisplaybreaks\\
&s_2 : (b_1, b_2, b_3, b_4, b_5, b_6, b_7, b_8, f, g) &\!\!\!\!\mapsto    \bigg( b_1, \frac{1}{b_8}, b_2b_3b_8, b_2b_4b_8, b_5, b_6, \nonumber\\
& &\lefteqn{b_7,\frac{1}{b_2}, f, \frac{b_2(b_8-f)g}{1-b_2f-fg+b_2b_8fg}\bigg)},\allowdisplaybreaks\notag\\
&s_3 : (b_1, b_2, b_3, b_4, b_5, b_6, b_7, b_8, f, g) &\!\!\!\!\mapsto  (b_1, b_2, b_3, b_4, b_8, b_6, b_7, b_5, f, g),\nonumber\\
&s_4 : (b_1, b_2, b_3, b_4, b_5, b_6, b_7, b_8, f, g)&\!\!\!\!\mapsto  (b_1, b_2, b_3, b_4, b_6, b_5, b_7, b_8, f, g),\label{weyle6}\\
&s_5 : (b_1, b_2, b_3, b_4, b_5, b_6, b_7, b_8, f, g)&\!\!\!\!\mapsto (b_1, b_2, b_3, b_4, b_5, b_7, b_6, b_8, f, g),\nonumber\\
&s_6 : (b_1, b_2, b_3, b_4, b_5, b_6, b_7, b_8, f, g)&\!\!\!\!\mapsto  \bigg(b_1b_3b_8, b_2b_3b_8, \frac{1}{b_8}, b_4, b_5, b_6, \nonumber\\
& &\lefteqn{b_7, \frac{1}{b_3}, \frac{f(b_8g-1)}{-b_3b_8+b_8g-fg+b_3b_8fg}, g\bigg)},\nonumber\allowdisplaybreaks\\
&\pi_1 : (b_1, b_2, b_3, b_4, b_5, b_6, b_7, b_8, f, g)&\!\!\!\!\mapsto \bigg( \frac{1}{b_4}, \frac{1}{b_3}, \frac{1}{b_2}, \frac{1}{b_1},\frac{1}{b_5},\frac{1}{b_6}, \frac{1}{b_7},\frac{1}{b_8}, g, f \bigg), \nonumber\\
&\pi_2 : (b_1, b_2, b_3, b_4, b_5, b_6, b_7, b_8, f, g)&\!\!\!\!\mapsto \bigg(\frac{1}{b_7}, \frac{1}{b_6}, \frac{1}{b_3b_5b_8}, \frac{1}{b_4b_5b_8}, b_8, \frac{1}{b_2}, \nonumber\\
& &\lefteqn{\frac{1}{b_1}, b_5, f, \frac{1-fg}{f+b_5b_8g-b_5fg-b_8fg}\bigg)}.\nonumber,
\end{eqnarray}
\begin{rem}\label{remark}
The actions $s_0$, $\cdots$, $s_6$, $\pi_1$, and $\pi_2$ are invariant under the following transformation of
\begin{equation}
(b_1, b_2, b_3, b_4, b_5, b_6, b_7, b_8; f, g) \mapsto(b_1/\lambda, b_2/\lambda, b_3/\lambda, b_4/\lambda, \lambda b_5, \lambda b_6, \lambda b_7, \lambda b_8; \lambda f, g/\lambda) \nonumber
\end{equation}
where $\lambda$ is any parameter.
\end{rem}
Then, we have the following lemma.
\begin{lemm}
The transformations (\ref{weyle6}) satisfy the fundamental relation of the affine Weyl group $W(E^{(1)}_6)$:

\begin{align*}
 s^2_i&=1 \quad(i=0, \cdots, 6),& &  & \pi_i^2&=1\quad(i=1, 2), \\ 
  (s_{i}s_{j})^3&=1 \quad (a_{ij}=-1), & &  &   (s_is_j)^2&=1 \quad (a_{ij}=0), \\
  (\pi_1\pi_2)^3&=1,& & \\
 s_i\pi_1&=\pi_1s_j  & & & ((i,j)&=(1, 0), (2, 6), (3, 3), (4, 4), (5, 5)),\\
  s_i\pi_2&=\pi_2s_j  & &  &((i,j)&=(0, 0), (1, 5), (2, 4), (3, 3), (6, 6)),
\end{align*}

\end{lemm}
\proof\\
Direct calculation. \hfill $\Box$

We put $q=b_1b_2b_3b_4b_5b_6b_7b_8$. It is well known that for any element $T$ of the translation subgroup of $W(E^{(1)}_6)$, the operation of $T$ is regarded as Painlev\'{e} equation. For example, put
\begin{equation}
T=r'r\label{T}
\end{equation}
where
\begin{align}
r'&=\pi_1 r \pi_1,\label{r'}\\
r&=\pi_2s_0s_5s_4s_5s_3s_4s_5s_2s_3s_4s_5s_1s_2s_3s_4s_5.
\end{align}
Then, the following theorem is obtained.

\begin{theo}\label{weylpade}
For $T  \in W(E^{(1)}_6)$ given the equation (\ref{T}), we have
\begin{align}
\bigg(\frac{1}{f\underline g}, \frac{1}{fg}\bigg)_1
&=\frac{\displaystyle{\bigg(\frac{b_5}{f}, \frac{b_6}{f}, \frac{b_7}{f}, \frac{b_8}{f}\bigg)_1}}{\bigg(\displaystyle{\frac{1}{b_1f}, \frac{1}{b_2f}}\bigg)_1}\label{weyl1}\\
\frac{(\overline f g, fg)_1}{f \overline f}
&=\frac{b_1b_2(b_5g, b_6g, b_7g, b_8g)_1}{q\bigg(\displaystyle{\frac{g}{b_3}, \frac{g}{b_4}}\bigg)_1}\label{weyl2}
\end{align}\allowdisplaybreaks
where $\underline{g}=T^{-1}(g)=r^{-1}(g)$, $\overline f=T(f)=r'(f)$.
Furthermore by operating $T$ to the parameters $(b_1, \cdots, b_8)$, they shift as
\begin{equation}\label{parametershift}
(b_1, b_2, b_3, b_4, b_5, b_6, b_7, b_8) \mapsto(b_1/q, b_2/q, qb_3, qb_4, b_5, b_6, b_7, b_8) 
\end{equation}
\end{theo}
\proof\\
Direct computation by using
\begin{eqnarray*}
s_2\bigg(\frac{1}{fg}\bigg)_1=\frac{\bigg(\displaystyle{\frac{1}{b_2g}}\bigg)_1}{\bigg(\displaystyle{\frac{b_8}{f}}\bigg)_1}\bigg(\frac{1}{fg}\bigg)_1, &&
\pi_2\bigg(\frac{1}{fg}\bigg)_1=\frac{\bigg(\displaystyle{\frac{b_5}{f}, \frac{b_8}{f}}\bigg)_1}{\bigg(\displaystyle{\frac{1}{fg}}\bigg)_1}.
\end{eqnarray*}
Then, we obtain the equation (\ref{weyl1}). The equation (\ref{weyl2}) is similar. 
And the relation (\ref{parametershift}) is derived by operating $T$ for the parameters $(b_1, b_2, b_3, b_4, b_5, b_6, b_7, b_8)$. \hfill $\Box$

\subsection{Painlev\'{e} equations along the various directions and B\"{a}cklund transformations}

By the Pad\'{e} method, we can construct $q$-Painlev\'{e} equations along the various direction $T_i$ of deformations. These equations are equivalent through some B\"{a}cklund transformations of variables $(f, g) \mapsto (f_i, g_i)$ where $f$ and $g$ are variables in section \ref{sece6}. The variable $f$ is defined by the equation (\ref{fa}) which does not depend on the direction. So, the variable $f$ does not depend on the direction. On the other hand, the variable $g$ is defined by the equation (\ref{g}) which depends on the direction. So, the variable $g$ depends on the direction in general. 

In this appendix, we will present the correspondence of variables $g_i$ by using B\"{a}cklund transformations.

For example, we take the following transformation $T_1$ in the Pad\'{e} method
\begin{equation}
T_1: a_1\mapsto qa_1, a_2\mapsto qa_2, a_3\mapsto qa_3, a_4\mapsto qa_4\nonumber
\end{equation}
Then, we get the following theorem
\begin{theo}\label{pade}
For the direction $T_1$, we get
\begin{align}
\bigg(\frac{1}{f_1 g_1}, \frac{q}{f_1\underline{g_1}}\bigg)_1
&=\frac{\bigg(\displaystyle{\frac{a_1}{f_1}, \frac{a_2}{f_1}, \frac{a_3}{f_1}, \frac{a_4}{f_1}}\bigg)_1}{\bigg(\displaystyle{\frac{1}{q^{N}f_1}, \frac{q}{f_1}}\bigg)_1}, \label{weylpainleveg}\\
\frac{\bigg(f_1 g_1, \displaystyle{\frac{\overline{f_1}g_1}{q}}\bigg)_1}{f_1\overline{f_1}}
&=\frac{q^{N-1}(a_1g_1, a_2g_1, a_3g_1, a_4g_1)_1}{(q^ma_1a_2g_1, q^n a_3a_4g_1)_1}\label{weylpainlevef}.
\end{align}
\end{theo}
\proof\\
By similar, calculation in section \ref{sece6}, we get the Lax equation $L_2$ and $L_3$ as
\begin{align}
L_2& : (a_3x, a_4x)_1y(x)-\bigg(\frac{x}{q^{N}}, \frac{x}{g_1}\bigg)_1y(qx)+c_1x(f_1x)_1\overline{y}(x)=0,\label{t1l2}\\
L_3&: c_2x\bigg(\frac{\overline{f_1}x}{q}\bigg)_1y(x)+(a_1 x, a_2 x)_1\overline{y}(x)-\bigg(x, \frac{x}{g_1}\bigg)_1\overline{y}\bigg(\frac{x}{q}\bigg)=0
\end{align}
where $c_1$ and $c_2$ are constants. Then we get the $q$-Painlev\'{e} equation (\ref{weylpainleveg}) and (\ref{weylpainlevef}).\hfill $\Box$

Here, we make a correspondence between the parameters in (\ref{weylpainleveg}), (\ref{weylpainlevef}) and the parameters in (\ref{weyl1}), (\ref{weyl2}). One of the possible choice is
\begin{align*}
b_1=q^{N}, &\quad b_2=1/q, & b_3=\displaystyle{\frac{1}{q^ma_1a_2}}, & \quad b_4=\displaystyle{\frac{1}{q^n a_3a_4}}, &
b_5=a_1, &\quad b_6=a_2, & b_7=a_3, &\quad  b_8=a_4.
\end{align*}
In the followings, we always identify the parameters $m$, $n$, $a_1$, $\cdots$, $a_4$ with $b_1$, $\cdots$, $b_8$. by this rule. In particular the direction $T_1$ above is $T$ in the equation (\ref{T}).

By using this correspondence, $T_1$ is rewritten as
\[
\begin{array}{cccccc}
T_1 :b_3\mapsto b_3/q^2, &  b_4 \mapsto b_4/q^2, &
b_5\mapsto q b_5, & b_6\mapsto qb_6, & b_7 \mapsto qb_7, & b_8 \mapsto qb_8.
\end{array}
\]
And the Painlev\'{e} equation (\ref{weylpainleveg}), (\ref{weylpainlevef}) are rewritten as
\begin{align}\label{painleveq}
\bigg(\frac{1}{f_1 g_1}, \frac{q}{f_1\underline{g_1}}\bigg)_1=\frac{\bigg(\displaystyle{\frac{b_5}{f_1}, \frac{b_6}{f_1}, \frac{b_7}{f_1}, \frac{b_8}{f_1}}\bigg)_1}{\bigg(\displaystyle{\frac{1}{b_1f_1}, \frac{1}{b_2f_1}}\bigg)_1}, &&
\frac{\bigg(f_1 g_1, \displaystyle{\frac{\overline{f_1}g_1}{q}}\bigg)_1}{f_1\overline{f_1}}=\frac{b_1b_2(b_5g_1, b_6g_1, b_7g_1, b_8g_1)_1}{\bigg(\displaystyle{\frac{g_1}{b_3}, \frac{g_1}{b_4}}\bigg)_1}.
\end{align}
By using transformation in remark \ref{remark} with $\underline{\lambda}/\lambda=1/q$, we transform the equation (\ref{painleveq}). Then, we get
\begin{eqnarray}
& &\bigg(\frac{1}{f_1 g_1}, \frac{1}{f_1\underline{g_1}}\bigg)_1=\frac{\bigg(\displaystyle{\frac{b_5}{f_1}, \frac{b_6}{f_1}, \frac{b_7}{f_1}, \frac{b_8}{f_1}}\bigg)_1}{\bigg(\displaystyle{\frac{1}{b_1f_1}, \frac{1}{b_2f_1}}\bigg)_1}, \label{painlevebasis}\\
& &\frac{(f_1 g_1, \overline{f_1}g_1)_1}{f_1\overline{f_1}}=\frac{b_1b_2(b_5g_1, b_6g_1, b_7g_1, b_8g_1)_1}{q\bigg(\displaystyle{\frac{g_1}{b_3}, \frac{g_1}{b_4}}\bigg)_1}.\label{painlevebasis2}
\end{eqnarray}

In the followings (\ref{type1}, $\cdots$, \ref{type4}), we will consider the four types of deformation directions $T_1, \cdots, T_4$ in terms of variables $g_1$, $\cdots$, $g_4$. 
 We will present correspondence between these variables $g_1$, $\cdots$, $g_4$ by using B\"{a}cklund transformations.
\subsubsection{$T_1 :b_3\mapsto b_3/q^2,\quad  b_4 \mapsto b_4/q^2,\quad
b_5\mapsto q b_5,\quad b_6\mapsto qb_6,\quad b_7 \mapsto qb_7,\quad b_8 \mapsto qb_8.$}\label{type1}
This type corresponds to the direction considered in theorem \ref{pade}. And $q$-Painlev\'{e} equation is the equation (\ref{painlevebasis}). We will use this direction as the reference direction. We compare the other three types with this reference direction $T_1$.

\subsubsection{$T_2: b_1 \mapsto b_1/q,\quad b_8 \mapsto q b_8$.}\label{type2}
This direction corresponds to $s_2T_1^{-1}s_2$. By Pad\'{e} method, the Lax pair for $T_2$ are
\begin{eqnarray}
& &L_2:\bigg(b_8x, \frac{x}{g_2}\bigg)_1y(x)-(b_5x, b_6x)_1y(qx)+c_1'x(f_1x)_1\overline{y}(x)=0, \label{t2l2}\\
& &L_3: c_2'x\bigg(\frac{\overline{f_1}x}{q}\bigg)_1y(x)+\bigg(\frac{x}{b_1}, \frac{x}{qg_2}\bigg)_1\overline{y}(x)-\bigg(\frac{b_7}{q}x, x\bigg)_1\overline{y}\bigg(\frac{x}{a}\bigg)=0
\end{eqnarray}
where $c_1'$ and $c_2'$ are constant. And the $q$-Painlev\'{e} equation is given by
\begin{equation}\label{type2painleve}
\bigg(\frac{1}{f_1 g_2}, \frac{1}{f_1\underline{g_2}}\bigg)_1=\frac{\bigg(\displaystyle{\frac{b_5}{f_1}, \frac{b_6}{f_1}, \frac{b_7}{f_1}, \frac{1}{b_2f_1}}\bigg)_1}{\bigg(\displaystyle{\frac{b_8}{f_1}, \frac{1}{b_1f_1}}\bigg)_1}.
\end{equation}
Now, we derive the relation between $g_1$ and $g_2$. We substitute $x=1/f_1$ for the equation (\ref{t1l2}) and the equation (\ref{t2l2}). And taking a ratio of these two equations, we get the following equation
\begin{equation}\label{type2g}
\bigg(\frac{1}{f_1g_1}, \frac{1}{f_1g_2}\bigg)_1=\frac{\bigg(\displaystyle{\frac{b_5}{f_1}, \frac{b_6}{f_1}, \frac{b_7}{f_1}}\bigg)_1}{\bigg(\displaystyle{\frac{1}{b_1f_1}}\bigg)_1}.
\end{equation}

In terms of B\"{a}cklund transformation, the relation (\ref{type2g}) can be written as
\begin{equation}
g_2=s_2T_1^{-1}(g_1).\nonumber
\end{equation}

\begin{rem}
The direction $T$ in section \ref{sece6} is related with this direction $T_2$ by exchange $b_6$ and $b_8$. And similarly as above, it follows that
\begin{equation}
\bigg(\frac{1}{f_1g_2}\bigg)_1=\frac{\displaystyle{\bigg(\frac{b_6}{f_1}\bigg)_1}}{\displaystyle{\bigg(\frac{b_8}{f_1}\bigg)_1}}\bigg(\frac{1}{f_1g}\bigg)_1.
\end{equation}
\end{rem}

\subsubsection{$T_3: b_1\mapsto b_1/q, \quad b_3 \mapsto b_3/q,\quad b_4 \mapsto b_4/q,\quad b_5\mapsto qb_5,\quad b_6 \mapsto qb_6, \quad b_7 \mapsto qb_7.\quad $}\allowdisplaybreaks
This direction corresponds to $ T_2=s_1s_2T_1s_2s_1$. By Pad\'{e} method, the Lax pair for $T_3$ is
\begin{eqnarray*}
& &\!\!\!\!\!\!L_2: (b_7x)_1y(x)-\bigg(\frac{x}{g_3}\bigg)_1y(qx)+c_1'x(f_1x)_1\overline{y}(x)=0,\\
& &\!\!\!\!\!\! L_3: c_2'x\bigg(\frac{b_5}{q}x, \frac{b_6}{q}x, \frac{\overline{f_1}x}{q}\bigg)_1y(x)+\bigg(b_5x, b_6x, \frac{b_7}{q}x, \frac{x}{b_1}\bigg)_1\overline{y}(x) -\bigg(b_7x, b_8x, x, \frac{x}{g_3}\bigg)_1\overline{y}\bigg(\frac{x}{q}\bigg)=0
\end{eqnarray*}
where $c_1'$ and $c_2'$ are constant. And the $q$-Painlev\'{e} equation is given by\allowdisplaybreaks
\begin{equation}\label{type3painleve}
\bigg(\frac{1}{f_1 g_3}, \frac{1}{f_1\underline{g_3}}\bigg)_1=\frac{\bigg(\displaystyle{\frac{b_5}{f_1}, \frac{b_6}{f_1}, \frac{b_7}{f_1}, \frac{1}{b_1f_1}}\bigg)_1}{\bigg(\displaystyle{\frac{b_8}{f_1}, \frac{1}{b_2f_1}}\bigg)_1}.
\end{equation}
Now, we derive the relation between $g_1$ and $g_3$. Similarly as \ref{type2}, we get the following equation
\begin{equation}\label{type3g}
\bigg(\frac{1}{f_1g_3}\bigg)_1=\frac{\bigg(\displaystyle{\frac{1}{b_1f_1}}\bigg)_1}{\bigg(\displaystyle{\frac{b_8}{f_1}}\bigg)_1}\bigg(\frac{1}{f_1g_1}\bigg)_1.
\end{equation}

In terms of B\"{a}cklund transformation, the relation (\ref{type3g}) can be written as
\begin{equation}
g_3=s_1s_2(g_1).\nonumber
\end{equation}

\subsubsection{$T_4: b_3\mapsto b_3/q, b_4 \mapsto b_4/q, b_5 \mapsto qb_5, b_8\mapsto qb_8$.}\label{type4}
This direction corresponds to $T_4=s_2s_1s_3s_2T_1^{-1}s_2s_3s_1s_2$. By Pad\'{e} method, the Lax pair for $T_4$ is
\begin{eqnarray*}
& &L_2: \bigg(b_8x, \frac{x}{g_4}\bigg)_1y(x)-\bigg(b_6x, \frac{x}{b_1}\bigg)_1y(qx)+c_1'x(f_1x)_1\overline{y}(x)=0,\\
& &L_3: c_2'x\bigg(\frac{\overline{f_1}x}{q}\bigg)_1y(x)+\bigg(b_5x, \frac{x}{qg_4}\bigg)_1\overline{y}(x)-\bigg(\frac{b_7}{q}x, x\bigg)_1\overline{y}\bigg(\frac{x}{q}\bigg)_1=0
\end{eqnarray*}
where $c_1'$ and $c_2'$ are constant. And the $q$-Painlev\'{e} equation is given by
\begin{equation}\label{type4painleve}
\bigg(\frac{1}{f_1 g_4}, \frac{1}{f_1\underline{g_4}}\bigg)_1=\frac{\bigg(\displaystyle{\frac{1}{b_1f_1},\frac{b_6}{f_1}, \frac{b_7}{f_1}, \frac{1}{b_2f_1}}\bigg)_1}{\bigg(\displaystyle{\frac{b_5}{f_1}, \frac{b_8}{f_1}}\bigg)_1}.
\end{equation}

Now, we derive the relation between $g_1$ and $g_4$. 
Similarly as \ref{type2}, we get the following equation
\begin{equation}\label{type4g}
\bigg(\frac{1}{f_1g_4}, \frac{1}{f_1g_1}\bigg)_1=\bigg(\displaystyle{\frac{b_6}{f_1}, \frac{b_7}{f_1}}\bigg)_1.
\end{equation}

In terms of B\"{a}cklund transformation, the relation (\ref{type4g}) can be written as
\begin{equation}
g_4=s_2s_1s_3s_2T_1^{-1}(g_1).\nonumber
\end{equation}

\end{document}